# Transmission of topological surface states through surface barriers

Jungpil Seo[1], Pedram Roushan[1], Haim Beidenkopf[1], Y. S. Hor[2], R. J. Cava[2], Ali Yazdani[1]

[1]*Joseph Henry Laboratories & Department of Physics, Princeton University, Princeton, New Jersey 08544, USA.*

[2]*Department of Chemistry, Princeton University, Princeton, New Jersey 08544, USA.*

**Topological surface states are a class of novel electronic states that are of potential interest in quantum computing or spintronic applications[1–7]. Unlike conventional two-dimensional electron states, these surface states are expected to be immune to localization and to overcome barriers caused by material imperfection[8–14]. Previous experiments have demonstrated that topological surface states do not backscatter between equal and opposite momentum states, owing to their chiral spin texture[15–18]. However, so far there is no evidence that these states in fact transmit through naturally occurring surface defects. Here we use a scanning tunnelling microscope to measure the transmission and reflection probabilities of topological surface states of antimony through naturally occurring crystalline steps separating atomic terraces. In contrast to non-topological surface states of common metals (copper, silver and gold)[19–23], which are either reflected or absorbed by atomic steps, we show that topological surface states of antimony penetrate such barriers with high probability. This demonstration of the extended nature of antimony's topological surface states suggests that such states may be useful for high current transmission even in the**

**presence of atomic scale irregularities—an electronic feature sought to efficiently interconnect nanoscale devices.**

Topological surface states occur on compounds for which strong spin–orbit interaction influences the bulk electronic band structure in a fundamental way[1–7]. The key experimental signatures of topological surface states are an odd number of crossings of their energy–momentum dispersion at the Fermi level and chiral spin texture that locks the surface electrons' momenta to their spin angular momenta[8–11,15–18]. Angle-resolved photoemission spectroscopy (ARPES) measurements have observed both signatures of topological surface states for a number of compounds with a bulk gap, the so-called topological insulators such as $Bi_{1-x}Sb_x$, $Bi_2Se_3$ and $Bi_2Te_3$ and on the (111) surface of the semimetal antimony[8–11] (Sb). The two features of topological surface states alter their response to crystalline imperfections, first by eliminating 180° backscattering, owing to scattering restriction imposed by their spin texture, and second by the virtue of the odd number of band crossings, which prevents the surface states from being fully gapped when perturbed. The combination of these features makes the topological surface states robust against localization; hence, they are predicted to wrap the surface of the topological bulk sample regardless of the presence of surface defects. Previous scanning tunnelling microscopy (STM) experiments have provided evidence for the absence of 180° backscattering of these surface states by examining the scattering from substitutional impurities and atomic step edges[15–18]. However, these experiments do not demonstrate whether topological surface states are in fact extended, as they do not probe the transmission properties of these states. In fact, it is possible that in the presence of spin–orbit interaction, 180° backscattering is forbidden but electronic states can still be localized by surface defects. An example of this situation is the spin-split surface states of Au(111) that become localized by scattering from atomic step edges[22,23]. Topological surface states must be immune to such localization. To test this unique feature of topological surface states, we have used STM to examine antimony's surface states and to measure their reflection by and transmission through atomic steps.

We have studied the (111) surface of antimony in a home-built cryogenic STM that operates at 4 K. Single crystals of antimony were cleaved *in situ* in ultrahigh vacuum to expose a pristine surface. The STM topographic image shows nanometre-sized atomic terraces that are separated by single atom-height steps along the $[1\bar{1}0]$ crystallographic direction (Fig. 1a). We have examined the spatial and energy dependence of the local density of states (LDOS) within the terraces by performing differential conductance (d*I*/d*V*) measurements along a line perpendicular to the step edges. In Fig. 1b, the spatially averaged spectroscopic measurements for terraces of different widths show peaks that signal the occurrence of quantized resonances. These spectroscopic features start at around –225 meV, which is the bottom of the topological surface state bands as measured by ARPES studies[8,18]. Resolving the spatial variation of the LDOS within each terrace (Fig. 1c), we find clear evidence for quantized resonances, with nearly equal spacing in energy (see also Supplementary Information). The quantized resonances and the standing-wave patterns in the LDOS are caused by scattering of antimony's topological surface states from the atomic step edges and can be analysed to obtain reflection properties of these steps.

We first demonstrate that the standing-wave patterns and the nearly linear quantized resonances in LDOS can be understood on the basis of the ARPES band structure of antimony's surface states and the spin selection rules that must be obeyed for scattering of topological surface states. The Fourier transform (Fig. 2b) of the d$I$/d$V$ measurement on the 390 Å-wide terrace (Fig. 2a) reveals two different quantized wavevectors, $q_A$ and $q_B$, for the standing-wave pattern of the LDOS. Unlike conventional confined states that involve the superposition of states with equal and opposite momenta[24–27] (k), the chiral spin texture of antimony eliminates the possibility of superposition of states with orthogonal spins. Consequently, the wavevectors (**q**) observed correspond to superposition between states with different values of **k** but similar spin states. From the schematic ARPES measurements of antimony's constant-energy contours (Fig. 2c) and their spin texture (Fig. 2d), we identify $q_A$ with scattering between adjacent hole pockets, which accordingly vanishes once the hole pockets overlap each other (around −110 meV in Fig. 2c). Similarly, $q_B$ corresponds to superposition between the electron and hole pockets with oppositely oriented momenta and parallel spins[18] (Fig. 2d). The scattering from the step edges and the geometric constraint set by the terrace width, $L$, result in quantization of these allowed superpositions of momentum states, such that $q_n = 2\pi n/L$, where n is positive integer and q is the scattering wavevector . Despite their quantization, the energy dependences of $q_A$ and $q_B$ are related to the dispersion measured by ARPES bands because $\mathbf{q}_n(E_n) = \mathbf{k}_i(E_n) - \mathbf{k}_f(E_n)$, where $k_i$, $k_f$ are the initial and final momentum states in an elastic scattering with energy $E_n$. The nearly linear energy dependences of $q_A$ and $q_B$ have slopes (~1.2 eV Å) that are in excellent agreement with ARPES measurements.

The energy widths of the quantized resonances in the LDOS contain information on the scattering properties of antimony's topological surface states. In general, energy-level broadening, $\Gamma$, is inversely proportional to the lifetime of a state, $\tau$: $\Gamma \propto \hbar/\tau$ ($\hbar$, Planck's constant divided by $2\pi$). Examining the broadening of the resonances as a function of energy for different terraces (Fig. 3), we find that $\Gamma(E) = \Gamma_L + \gamma(E - E_F)^2$, where $E_F$ is the Fermi energy (with $\hbar/\gamma = 0.37 \pm 0.01$ fs eV$^2$, for terraces where level broadening is smaller than level spacing). This functional form illustrates the role of electron–electron scattering in the decoherence of topological surface states, similar to the case in other Fermi liquids[20]. However, $\Gamma_L$, the residual finite resonance width at $E_F$, characterizes the degree to which step edges can reflect antimony's topological surface states in the same spin channel. Using a simple model of energy broadening based on a double-barrier potential, we find that the energy broadening corresponds to a 42 ± 4% reflection probability for this process (Supplementary Information). This reflectivity is much lower than those reported for free-electron-like surface states on copper, silver or gold from STM and other measurements; however, it can be a result of scattering of the surface states into the bulk states rather than their transmission. Indeed, previous studies show that non-topological noble-metal surface states are absorbed in the bulk states with 30–50% probability without transmission when scattered by step edges[19–23].

Finite transmission through the atomic steps edges should give rise to coupling between electronic states on adjacent terraces. To search for evidence of such effects, we examine a configuration of atomic steps consisting of a narrow (110 Å) and a wide (2,500 Å) terrace, which resembles a nanoscale Fabry–Pérot resonator, as shown in Fig. 4. The narrow terrace shows the signature of quantized resonances according to the spin scattering rules described above (Fig. 4b). On the wide terrace, which is effectively semi-infinite, the LDOS oscillation forms an almost continuous pattern (Fig. 4c) with diverging wavelength towards the bottom of the surface state bands. The standing-wave patterns on the wide terrace also obey the chiral spin texture and are made up of the two wavevectors $q_A$ and $q_B$ described above, as illustrated in the Fourier transform in Fig. 4d. Any finite transmission through the step edges results in resonant tunnelling of surface state electrons from the wide terrace through the electron states of the narrow terrace.

Evidence for resonant tunnelling of the topological surface states can be seen both in the structure of energy-resolved modulations of the LDOS and in its Fourier transform (Fig. 4). At energies corresponding to the quantized resonances of the narrow terrace, there are clear suppressions of the modulation in the LDOS on the wide terrace. We illustrate this in Fig. 4e by plotting the Fourier intensity of the $q_B$ peak as a function of energy. Once the overall background changes in the LDOS (Fig. 4e, dashed green line) are removed, the intensity of the $q_B$ Fourier component displays characteristic signs of Fabry–Pérot resonant tunnelling (Fig. 4f) in which the surface state electrons pass through the narrow terrace without reflection. Deviation from perfect transmission is, however, a signature of scattering of the surface states into the antimony bulk states, from which we can deduce that the probability of this process is 23 ± 7% (Supplementary Information). Similarly, the width of the Fabry–Pérot resonance (Fig. 4f) can be used to extract the transmission and reflection probabilities and to show that they are 35 ± 3% and 42 ± 4%, respectively. In contrast, STM data on Fabry–Pérot structures for non-topological surface states of Ag(111) show no evidence of resonant tunnelling and can be explained in detail by considering only reflections and absorption of the surface states at the step edges[21].

The remarkable finding that topological surface states are as likely to be transmitted through an atomic step edge as reflected by it demonstrates an important difference between topological states and typical surface states in other materials. The transmissibility illustrates the extended nature of topological surface states even in the presence of strong surface barriers. The nearly equal probabilities of reflection and transmission at the step edges in antimony can also be understood on the basis of the topological surface state's band structure, which includes both forwards- and backwards-moving momenta with the same spin orientation (Fig. 2). The observation of high transmission also suggests excellent transport of topological surface states despite boundaries that would suppress the ability of other surface states to carry current. Such a feature may find applications in nanoscale devices, where there is a need for high surface conductivity because the surface-to-volume ratio is large.

Our experiments also show that nanoscale structures can be used to explore the properties of topological surface states both for fundamental studies and to evaluate their potential for device applications. Clearly, an extension of these studies to materials with a bulk gap and simpler surface band structure (such as in $Bi_2Te_3$ or $Bi_2Se_3$) would allow further examination of these states without the complication of coupling to the bulk states. However, antimony's surface band structure allows STM characterization of both reflection and transmission scattering rates within the same spin channel, which would not be possible for a single Dirac-cone surface band structure. Nevertheless, our demonstrations of the transmission of topological surface states through barriers that fully reflect other surface states suggest an extraordinary insensitivity of transport through these states to the geometrical shape of the surface.

**Acknowledgements** We gratefully acknowledge discussions with B. A. Bernevig, B. Boyanov, M. Z. Hasan and N. P. Ong. This work was supported by grants from the NSF-MRSEC programme through the Princeton Centre for Complex Materials, the ARO, the DOE, the NSF-DMR and the W. M. Keck Foundation. P.R. acknowledges support by a NSF graduate fellowship.

**Author Contribution** Y.S.H and R.J.C. carried out the growth of the single crystals and characterized them; STM measurement and data analysis were done by J.S., P.R., H.B., and A.Y. All authors discussed the result and contributed to the writing the manuscript.

**Author Information** Correspondence and requests for materials should be addressed to A.Y. (yazdani@princeton.edu).

## METHODS SUMMARY

### Crystal growth

We cleaved a single crystal of antimony from a boule grown by the modified Bridgman method. Before the growth, high-purity antimony (99.999%) was sealed in a vacuum quartz tube and heated to 700 °C. The crystal growth process took about a day at a slow rate of cooling from 700 to 500 °C. We then kept the boule at 500 °C for annealing over five days. The boule was then furnace-cooled down to 25 °C.

### STM measurement

We performed the experiment using a home-built cryogenic STM that operates at 4 K in ultrahigh vacuum. The typical sample size used was about 2 mm × 2 mm × 1 mm. The sample was cleaved *in situ* at room temperature in ultrahigh vacuum before being measured at low temperature. The cleaved sample revealed a (111) surface orientation due to the rhombohedral antimony crystal structure. A mechanically sharpened platinum–iridium alloy wire was used as a STM tip, and the quality of the tip was checked by imaging the surface states of a copper sample. The STM topography was measured in a constant-current mode, and d$I$/d$V$ signals were obtained by a standard lock-in technique with frequency $f$ = 757 Hz and 1-mV root-mean-squared modulation.

**Figure 1 The topological surface states of Sb(111) on atomic terraces. a**, STM topographic image ($V_{bias}$ = 1 V, $I$ = 8 pA) of a 2,500 Å-by-1,250 Å area showing terraces of various widths separated by 3.7 Å-high single atomic steps. **b**, The d$I$/d$V$ measurement averaged over each terrace shows quantized peaks. The spectra are offset vertically for clarity. **c**, Spatially and energetically resolved d$I$/d$V$ measurements taken along the dotted arrow in a demonstrate the interference in space and the quantization in energy.

**Figure 2 Allowed scattering wavevectors and their quantization. a**, The d$I$/d$V$ measurement on a 390 Å-wide terrace. **b**, Energy-resolved Fourier transform of the spatial modulation of the data in a reveals the quantization of scattering wavevectors $q_A$ and $q_B$. **c**, The dispersions of $q_A$ and $q_B$ match the dispersion of the surface bands as measured by ARPES along the high-symmetry directions (solid lines) and extend it above the Fermi level (dotted lines). **d**, Contours of constant energy of the antimony surface state (ARPES measurements from ref. 8). The contours consist of a central electron pocket and six surrounding hole pockets. The coloured arrows represent spin texture of the surface state. The scattering wavevectors $q_A$ and $q_B$ indicate allowed scattering processes. $\bar{\Gamma}$, $\bar{M}$ and $\bar{K}$, are the high symmetry points of the first Brillion

zone, with $\bar{\Gamma}$ located at the centre of the zone, and $\bar{M}$ in the middle of each side of this six-fold symmetric zone, and $\bar{K}$ at the vertex.

**Figure 3 Lifetime and leakage of quantized quasiparticles.** Each resonant peak is fitted to a Lorentzian function to yield the full-width at half-maximum of the quantized energy peaks $\Gamma$ (the inset). The plot shows the energy dependence of $\Gamma$ for two different terraces. The dashed lines are parabolic fits, and $\Gamma_L$ is the peak broadening at the Fermi energy. The error bars are from the fitting process of the resonant peaks and are mainly due to the measurement resolution.

**Figure 4 Resonant tunnelling between adjacent terraces. a**, STM topographic image of a narrow terrace ($L$ = 110 Å) and an adjacent wide terrace ($L$ = 2,500 Å). Only part of the wide terrace is displayed. The colour scale shows the height variations in the topographic image. **b**, On the narrow terrace, the d$I$/d$V$ measurement shows the quantization of the energy levels. **c**, The d$I$/d$V$ measurement on the wide terrace shows sudden phase shifts and suppressions of the d$I$/d$V$ intensities at around +5 meV and −70 meV due to resonant tunnelling. The averaged background conductance at each energy has been subtracted. **d**, The Fourier transform of the spatial modulation of the data in c. The grey markers indicate suppression of modulation intensities. **e**, Spectral weight of the scattering wavevector $\mathbf{q_B}$ as extracted from **d**. The dashed line corresponds to the background spectral weight in the absence of resonance tunneling. **f**, Spectral weight (circles) normalized by the background. The best fit (red line) yields ~42% reflection, ~35% transmission and ~23% bulk absorption from the scattering

process at the boundary. The blue line, which is the fit based on a model without bulk absorption, is displayed for comparison.

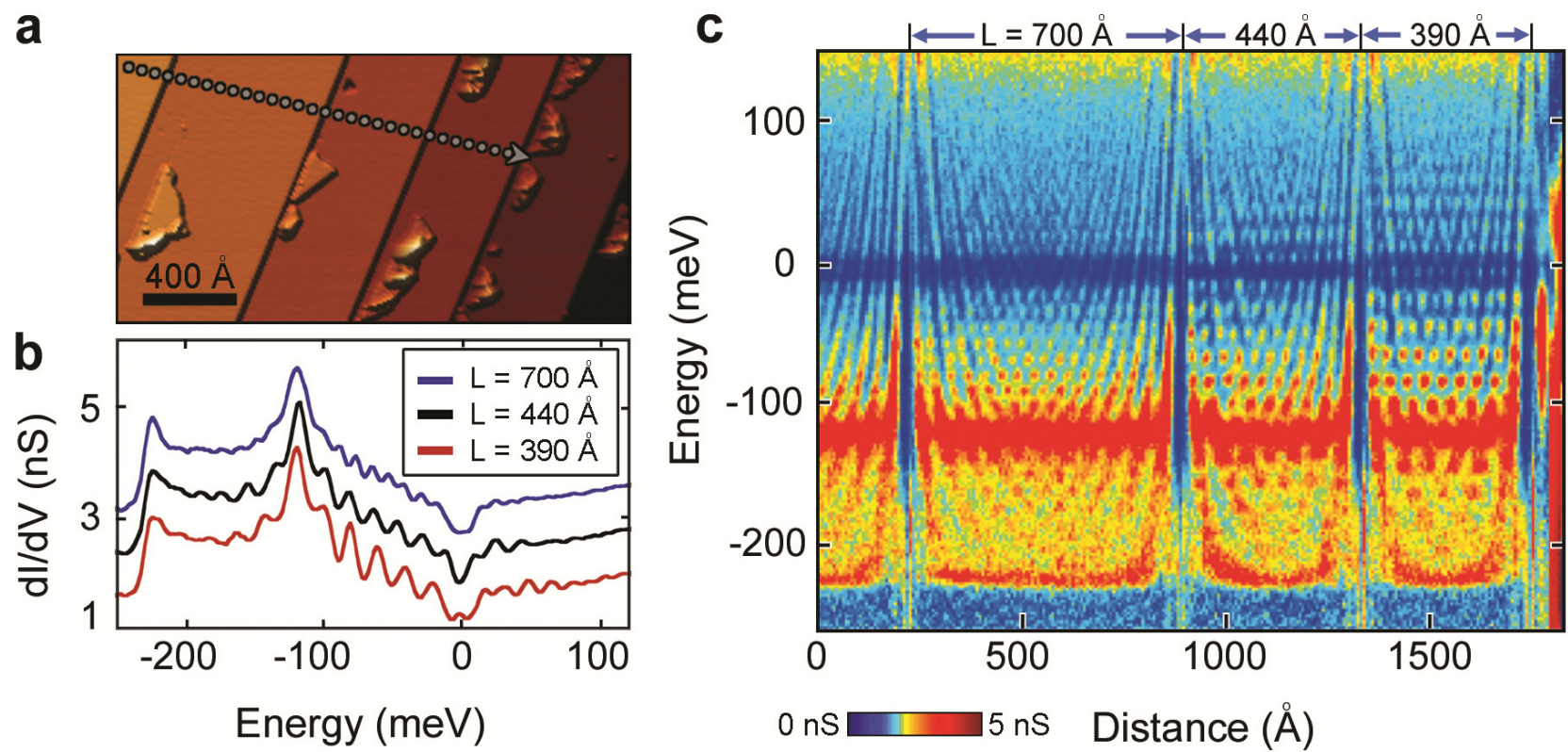

Figure 1

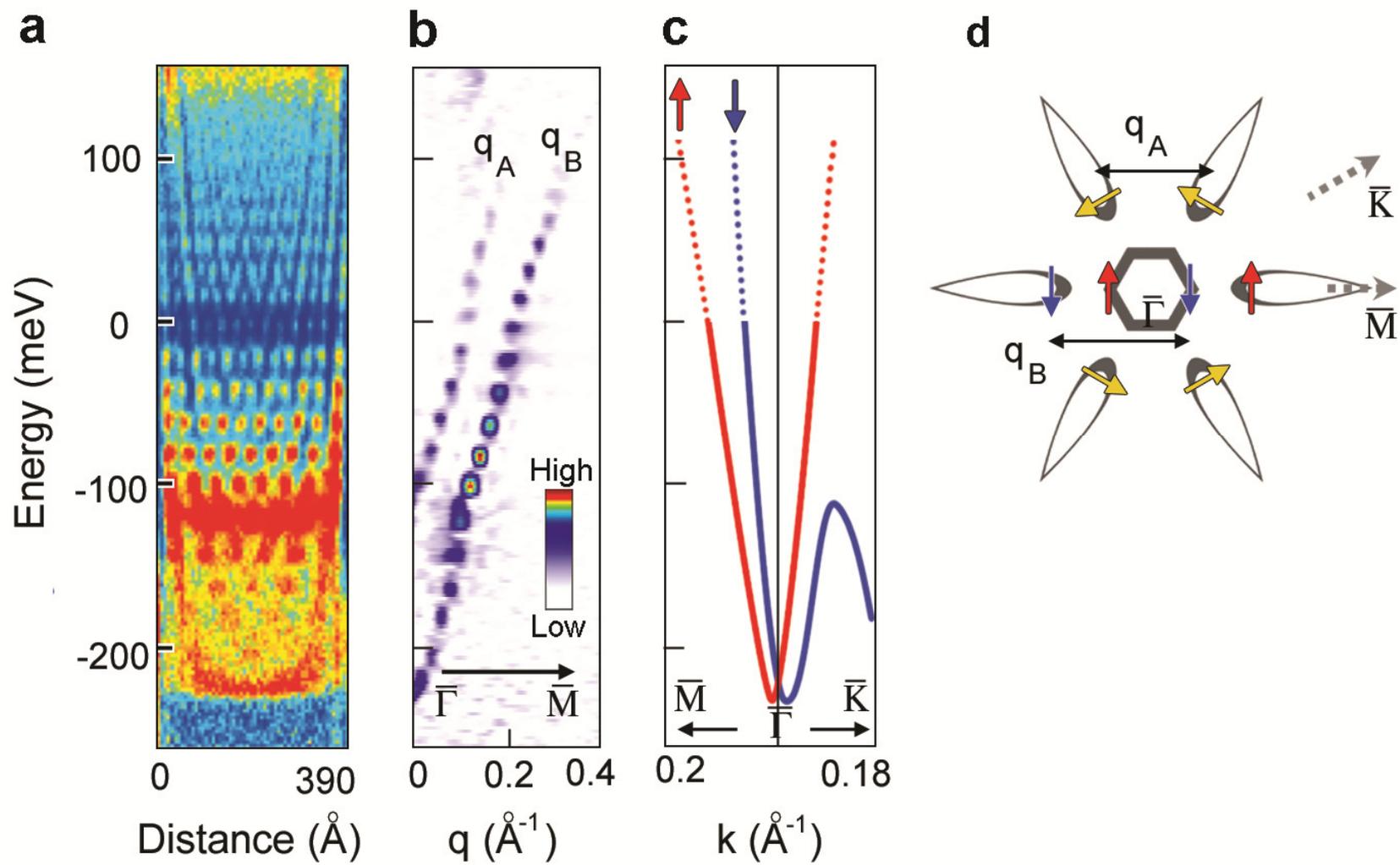

Figure 2

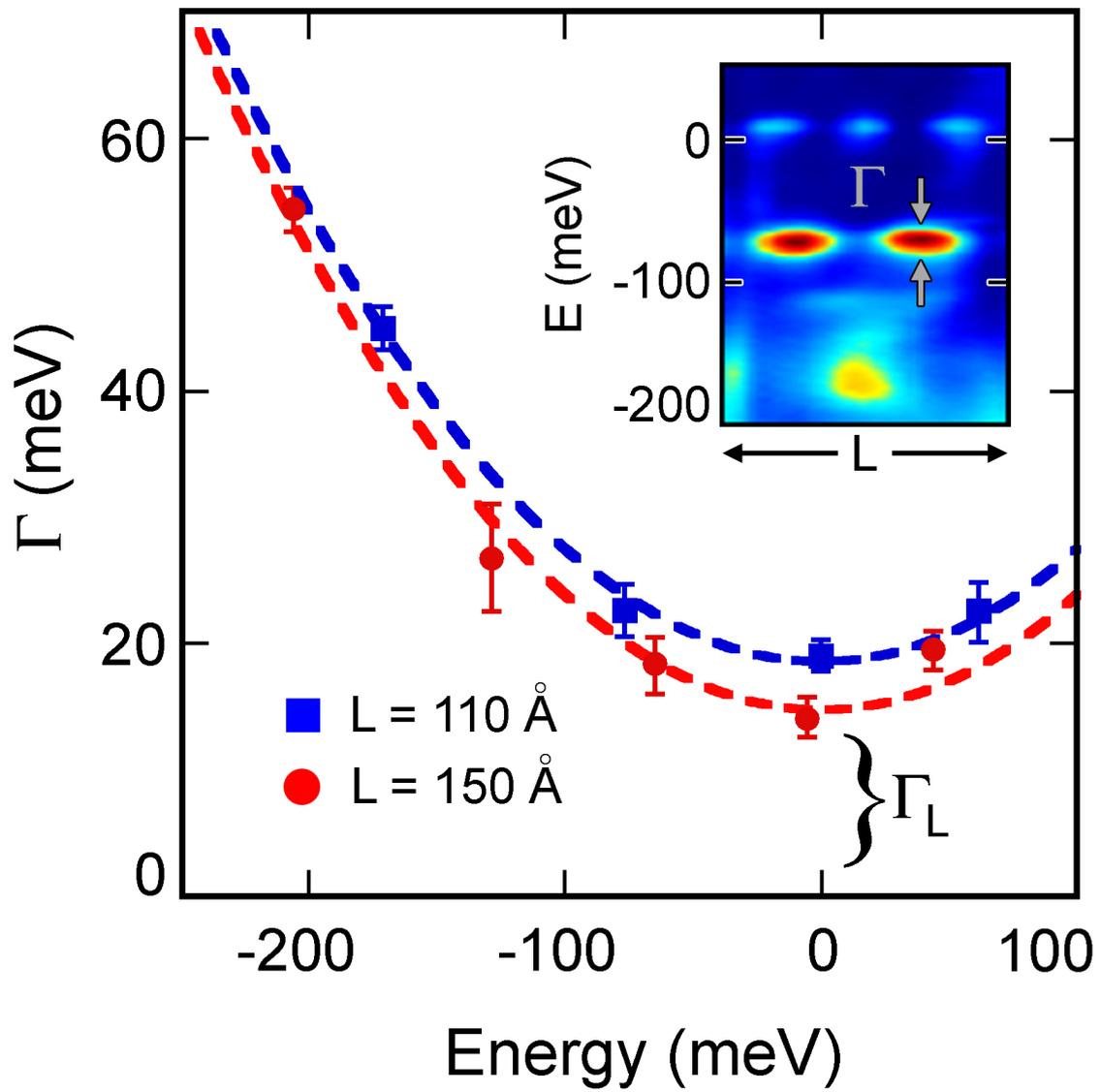

Figure 3

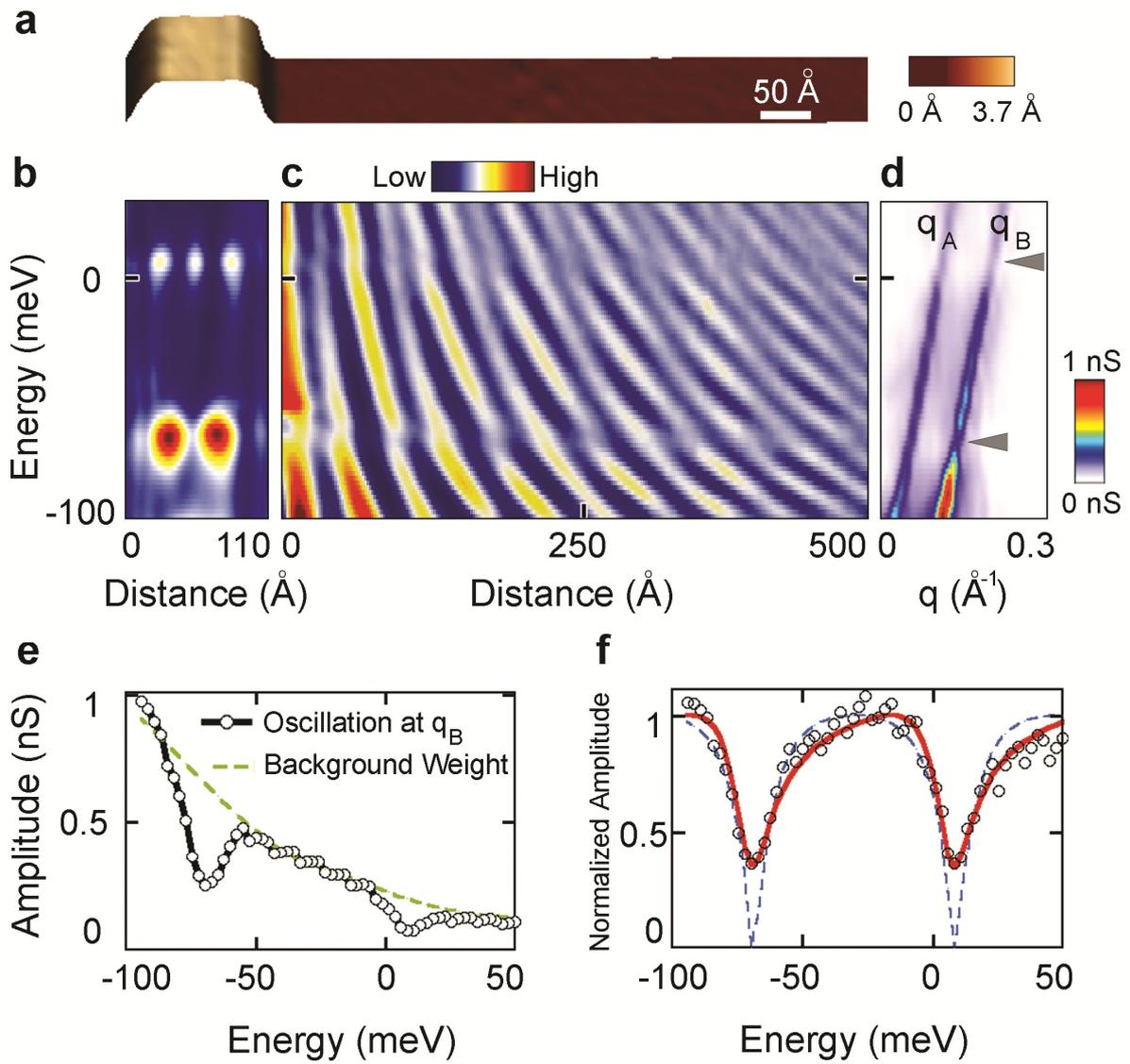

Figure 4

Supplementary Information Section

## 1. Quantization of linear band structure of Sb

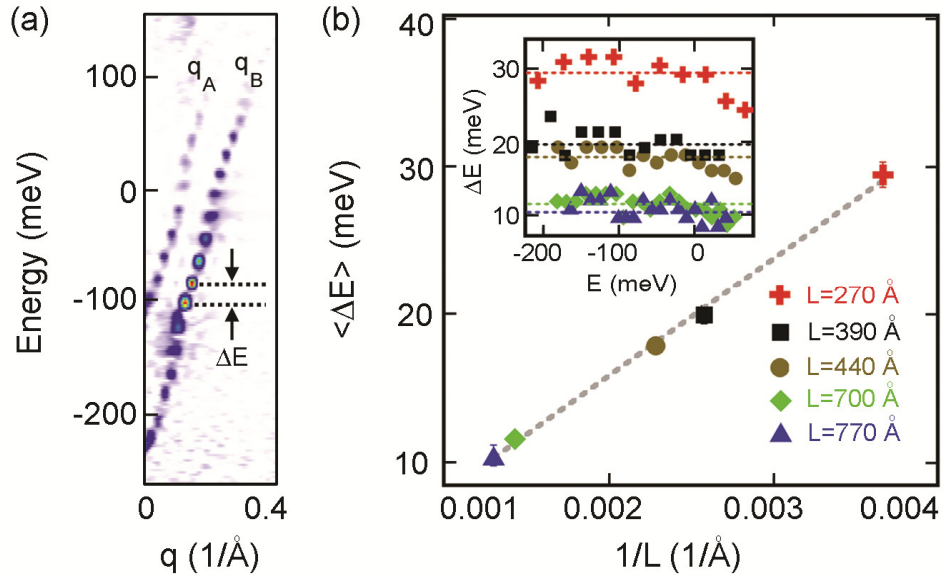

**Figure S1. Quantization of linear band structure of Sb.** (**a**) Energy resolved Fourier transform of a 390 Å wide terrace (From Fig. 2 in the main text). ΔE is defined with the spacing between adjacent quantized levels. (**b**) The energy level separations plotted for different terraces show inverse proportionality to the terrace width. The inset is a plot of the energy spacing between quantized energy levels at different energies, and the dashed lines are their average values. This plot is made from $q_B$ dispersion. $q_A$ dispersion gives similar results.

We measured the energy level spacing between adjacent quantized levels (ΔE) for $q_B$ dispersion to study the nature of Sb surface states (Fig. S1 (a)). In the inset of Fig. S1 (b), we plot ΔE as a function of energy for different terrace sizes, where for the entire range of measurements ΔE is essentially independent of energy. The average values of ΔE for each terrace, shown in Fig. S1 (b), illustrate that the level spacing is also inversely proportional to terrace width, which is also consistent with linear dispersion of the Sb band structure, $E \propto k$.

## 2. The measurement of the reflection probability in a Sb terrace

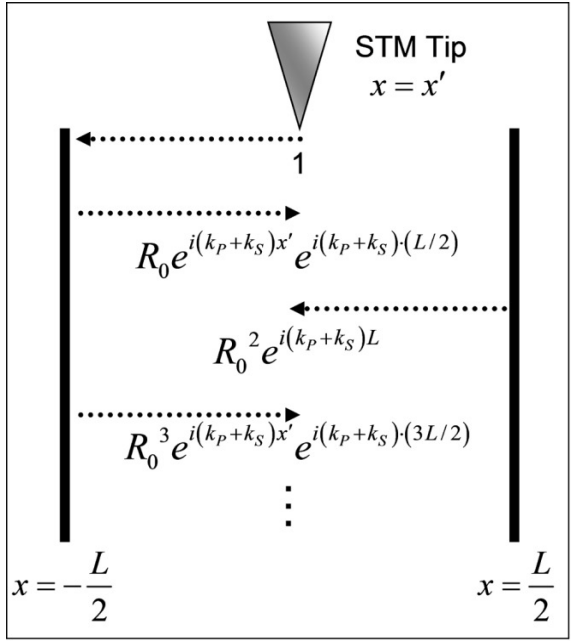

**Figure S2. Multiple scatterings in a quantum box.** The left propagating wave has momentum $k_P$ and the right propagating wave has $k_S$, where $P$ and $S$ refer to specific points of the Fermi contour (Fig. S3 (a)).

In order to quantify reflection probability of quasi-particles in a Sb terrace, we used a simple model based on multiple scatterings [*S1*]. In this approach, the electrons injected from a scanning tunneling microscope (STM) tip return to the position of the tip after multiple scatterings. The wavefunction amplitude is phase superposition from all possible scattering paths.

When the STM tip is positioned at $x = x'$ on the Sb terrace, the electrons from the tip propagate to both atomic boundaries. For the electron that initially travels to the left wall (Fig. 2S), the wavefunction amplitude $\psi_1$ at $x = x'$ is,

$$\psi_1 = 1 + e^{ik_P(x'+L/2)} \cdot R_0 \cdot e^{ik_S(x'+L/2)} + e^{ik_P(x'+L/2)} \cdot R_0 \cdot e^{ik_S L} \cdot R_0 \cdot e^{ik_P(L/2-x')} +$$
$$e^{ik_P(x'+L/2)} \cdot R_0 \cdot e^{ik_S L} \cdot R_0 \cdot e^{ik_P L} \cdot R_0 \cdot e^{ik_S(x'+L/2)} + \cdots$$
$$= \frac{1}{1-R_0^2 e^{i(k_P+k_S)L}} + \frac{R_0 e^{i(k_P+k_S)\cdot(L/2)}}{1-R_0^2 e^{i(k_P+k_S)L}} e^{i(k_P+k_S)x'},$$

where $L$ is the width of the terrace, and the reflection coefficient $R_0$ is a complex number to account for phase shift during the scattering process at the boundary. $k_P$ is the wavevector for the left going waves and $k_S$ is for the right going one. $P$ and $S$ are two high density points of the Fermi contours which have the same spin state (see Fig. 3S (a), the result would be identical if the other spin states $Q$ and $T$ were considered).

Similarly, the wavefunction amplitude $\psi_2$ for the electron that initially travels to the right wall is

$$\psi_2 = \frac{R_0 e^{i(k_P+k_S)\cdot(L/2)}}{1-R_0^2 e^{i(k_P+k_S)L}} e^{-i(k_P+k_S)x'} + \frac{1}{1-R_0^2 e^{i(k_P+k_S)L}}.$$

The recorded local density of states (LDOS) by STM is related to both of these wave amplitudes and is given by

$$LDOS(k,x') = \frac{1}{N}\left(|\psi_1+\psi_2|^2\right)$$
$$= \frac{1}{N}\left(\frac{4+4r^2\cos^2\left([k_P+k_S]x'\right)+8r\cos\left([k_P+k_S]x'\right)\cos\left(\theta_r+[k_P+k_S]\cdot(L/2)\right)}{1+r^4-2r^2\cos\left(2\theta_r+[k_P+k_S]L\right)}\right), \quad \cdots(1)$$

where $N$ is the normalization factor, and the reflection coefficient is expressed as $R_0 = re^{i\theta_r}$ ($r$, $\theta_r$ : real number).

The resonant condition is satisfied when $\cos\left(2\theta_r+[k_P(E)+k_S(E)]L\right)$ is 1. The full width at half maximum (FWHM) of the resonance peak is obtained from the Taylor expansion of the denominator of Eq.(1):

$$\Gamma_L = 2\sqrt{\frac{c^2\left(1-|R_0|^2\right)^2}{L^2|R_0|^2}}, \quad \cdots(2)$$

where $c = \left(c_P^{-1} + c_S^{-1}\right)^{-1}$, $E = c_P k_P$, and $E = c_S k_S$. The coefficients $c_P$ and $c_S$ are the slopes in the linear dispersion of the surface bands. They are known to be $c_p \cong 3.8$ V·Å and $c_S \cong 1.7$ V·Å from ARPES measurement [S2], and resulting c value $\approx 1.2$ V·Å is consistent with our finding in Fig. S1.

The values of $\Gamma_L$, are 19.4 mV and 14.3 mV for two different terraces of L=110 Å and L=150 Å, respectively (Fig. 3 of the main text). From Eq. (2) these values yields $|R_0|^2 = 0.42\pm0.04$. Figure S3 (b) and (c) show good agreement between the STM data and the simulated image by Eq. (1).

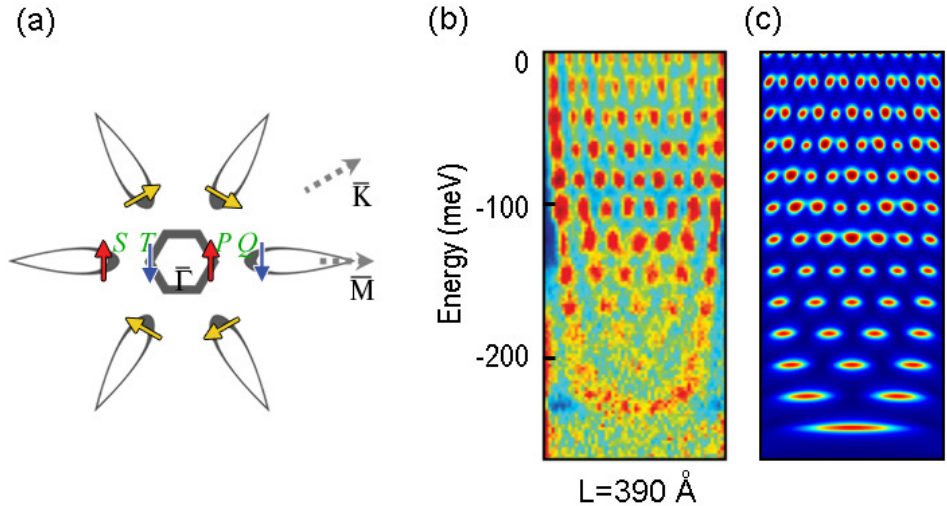

**Figure S3. Quantized topological states in a box.** (**a**) Contours of constant energy on the Sb surface. (**b**) dI/dV modulation in STM measurement. Averaged background conductance is subtracted at each energy to enhance the contrast. (**c**) The calculated image by Eq. (1).

## 3. Transmission and absorption probability measured from resonant tunneling

When a wide terrace exists adjacent to a narrow terrace, the wavefunction of the electron in the wide terrace can be modeled by considering the scatterings from both edges of the narrow terrace (Fig. S4). When the STM tip is positioned at $x = x'$ away from the near edge of the narrow terrace, the total reflection amplitude $A$ after multiple scatterings is,

$$A e^{i(k_P+k_S)x'} = e^{ik_P x'} R_0 e^{ik_S x'} + e^{ik_P x'} T_0 e^{ik_P L} R_0 e^{ik_S L} T_0 e^{ik_S x'} + e^{ik_P x'} T_0 e^{ik_P L} R_0 e^{ik_S L} R_0 e^{ik_P L} R_0 e^{ik_S L} T_0 e^{ik_S x'} + \cdots$$

$$= \left[ R_0 + \frac{T_0^2 R_0 e^{i(k_P+k_S)L}}{1 - R_0^2 e^{i(k_P+k_S)L}} \right] e^{i(k_P+k_S)x'},$$

where the reflection coefficient $R_0$ and transmission coefficient $T_0$ are complex numbers that represent the phase shift during the scattering process. The total wavefunction amplitude at the position of the STM tip is $1 + A e^{i(k_P+k_S)x'}$.

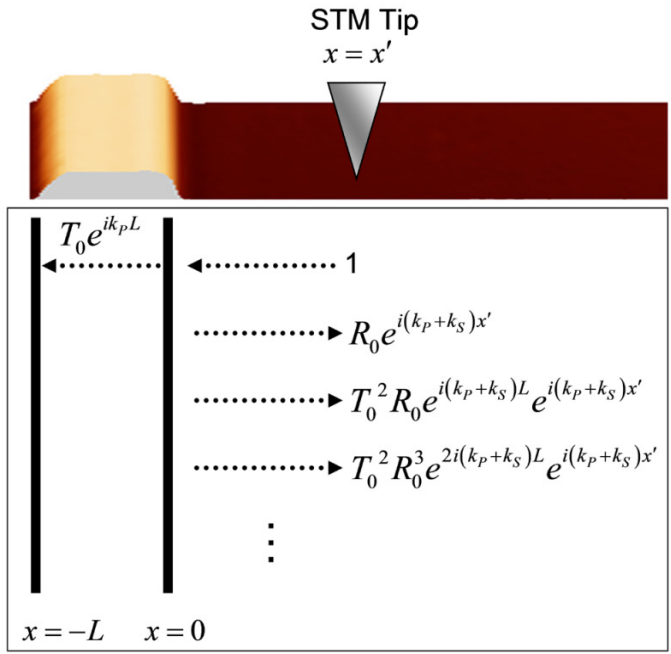

**Figure S4. Multiple scatterings from two adjacent atomic step edges.** The total wavefunction amplitude at the tip position is superposition of electron's phases resulting from all possible scattering paths.

The LDOS is expressed by

$$LDOS(k,x') = \frac{1}{N}\left(1+Ae^{i(k_P+k_S)x'}\right)\left(1+Ae^{i(k_P+k_S)x'}\right)^*$$
$$= \frac{1}{N}\left[1+|A|^2+(A+A^*)\cos(k_P+k_S)x'+i(A-A^*)\sin(k_P+k_S)x'\right],$$

where $N$ is the normalization factor. The pattern shows a spatial oscillation with wavenumber of $(k_P+k_S)$ and amplitude of

$$f(T_0,R_0) = \frac{1}{N}\sqrt{(A+A^*)^2+[i(A-A^*)]^2} = \frac{2}{N}|A|.$$

To account for absorption by the bulk, we introduce a factor, $\alpha$, and modify the wavefunction continuity and particle conservation relations accordingly [S3]:

$$1+R_0 = T_0$$
$$|R_0|^2+|T_0|^2 = 1-\alpha.$$

By setting $R_0 = re^{i\theta_r}$ and $T_0 = te^{i\theta_t}$, $\theta_r, \theta_t$ can be expressed in terms of $\alpha, r, t$.

$$r^2+t^2 = 1-\alpha, \quad \theta_r(\alpha,r) = \arccos\left(-\frac{\alpha}{2r}-r\right), \quad \theta_t(\alpha,t) = \arccos\left(\frac{\alpha}{2t}+t\right).$$

Hence, the oscillation amplitude can be rewritten as

$$f(t,r) = \frac{2}{N}\left(r^2+\frac{r^2t^4-2r^4t^2\cos(2\theta_t-2\theta_r)+2r^2t^2\cos(2\theta_t+2L\cdot E/c)}{1+r^4-2r^2\cos(2\theta_r+2L\cdot E/c)}\right)^{1/2}, \quad \cdots(3)$$

where $c = (c_P^{-1}+c_S^{-1})^{-1}$, $E = c_P k_P$, and $E = c_S k_S$.

From residual peak broadening ($\Gamma_L$ in Fig. 3 of the main text, and the fitting procedure described in section 2), $r^2$ is found to be 0.42±0.04. Substituting this value in Eq. (3) allows fitting the data in Fig. 4f of the main text with two free parameters N and $\alpha$, yielding $\alpha$=0.23±0.07. The transmission probability is accordingly set to $t^2$=0.35±0.03. Fig. S5 shows the STM data and the simulated image from Eq. (3) using the obtained values. In addition to matching the oscillation pattern, the qualitative agreement of phase shifts at the resonant energies reassures the obtained values for *t, r,* and α.

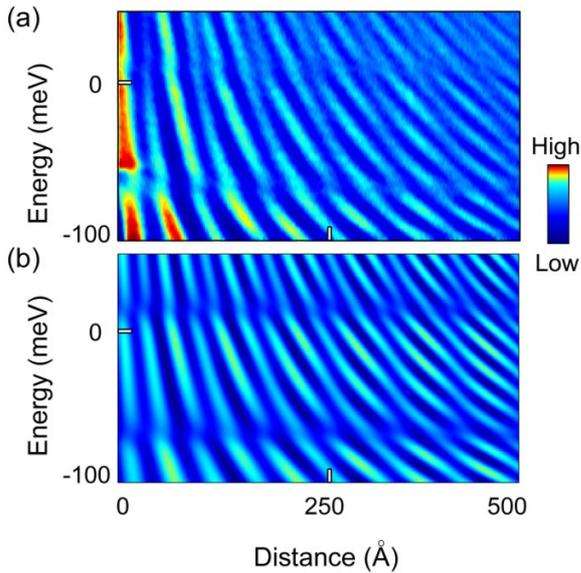

**Figure S5. Modeling the modulations in the wide terrace.** (**a**) dI/dV measurements on a wide terrace next to a narrow terrace. Averaged background conductance has been subtracted. (**b**) The image calculated by Eq. (3).

### References

S1. Heller, E. J., Crommie, M. F., Lutz, C. P. & Eigler, D. M.  Scattering and absorption of surface electron waves in quantum corrals. *Nature* **369**, 464–466 (1994).

S2. Hsieh, D. *et al.* Observation of unconventional quantum spin textures in topological insulators. *Science* **323**, 919–922 (2009).

S3. García-Calderón, G. & Chaos-Cador, L. Theroy of coherent and incoherent processes in quantum corrals. *Phys. Rev. B* **73**, 195405 (2006)